\documentstyle[11pt,newpasp,twoside]{article}
\def\edcomment#1{\iffalse\marginpar{\raggedright\sl#1\/}\else\relax\fi}
\reversemarginpar

\begin{document}
\title{Observations of YSO Circumstellar Discs}
 \author{D. Ward-Thompson}
\affil{Dept of Physics \& Astronomy,
Cardiff University, PO Box 913, Cardiff}
\author{P. Andr\'e}
\affil{CEA,Service d'Astrophysique, 
C.E. Saclay, Gif-sur-Yvette, France}
\author{O. P. Lay}
\affil{Jet Propulsion Lab, 4800 Oak Grove 
Drive, Pasadena, California}

\begin{abstract}
We present the results of short baseline interferometry observations
at submillimetre wavelengths, using a two-element interferometer
comprising the JCMT/CSO, of circumstellar discs around young
YSOs. We model data for the Class 0 protostar IRAS03282
and the Class I protostar L1709B with discs
and compare them with previously published results.
We find evidence for the ratio of the disc to envelope mass
to increase as the objects evolve.
\end{abstract}

\section{Introduction}

Circumstellar discs now appear to be an ubiquitous phenomenon
around newly-formed stars. One of the most direct ways to study discs is
to use submm interferometry. For optically thin emission,
the submm flux density is proportional to the mass of the
circumstellar disc. 
A number of young stars have now been studied with submm interferometers,
including that involving the linking of the JCMT \& CSO
to perform short-baseline interferometry (SBI).

\begin{figure}
\setlength{\unitlength}{1mm}
\begin{picture}(50,50)
\includegraphics{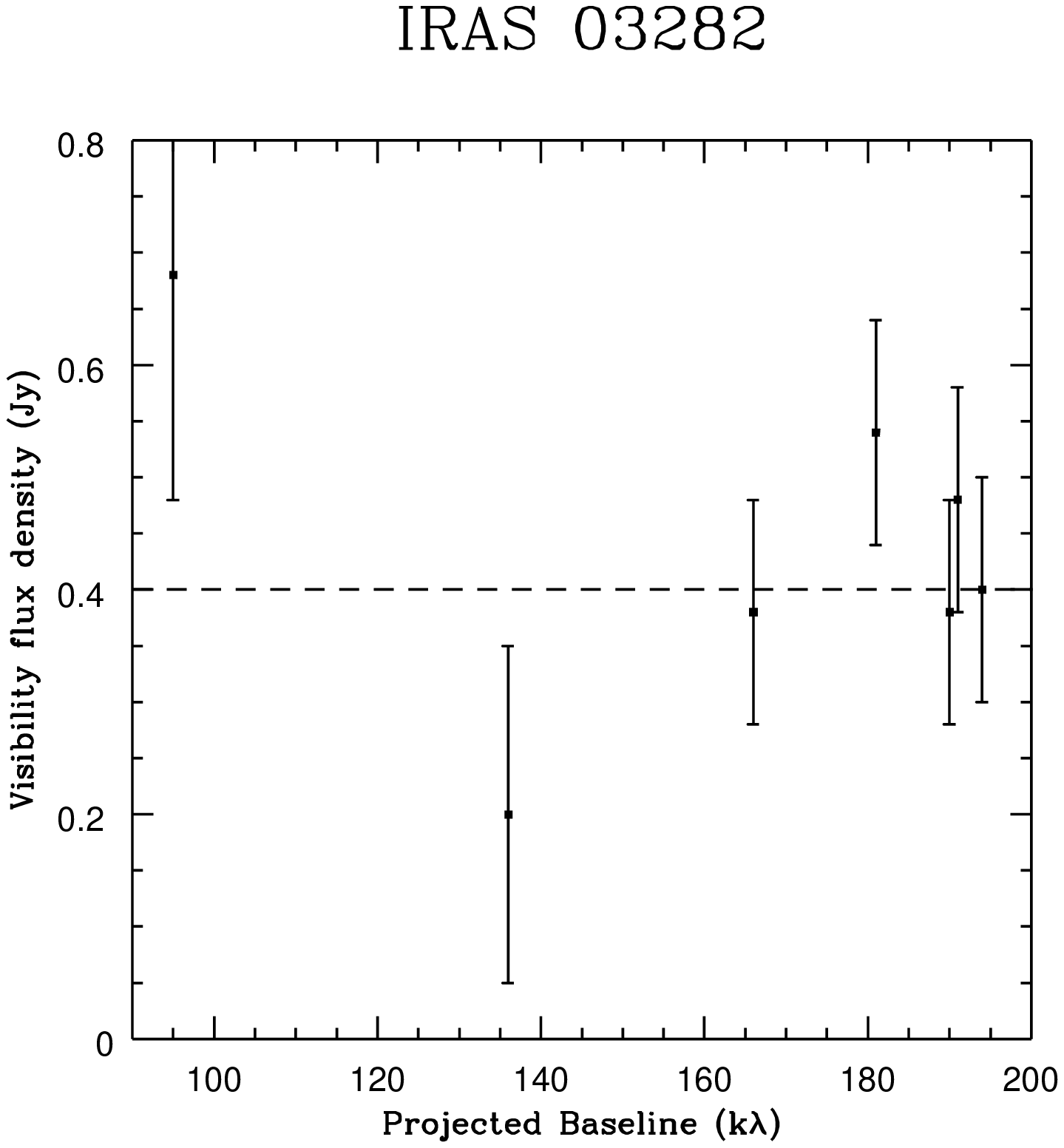}
\includegraphics{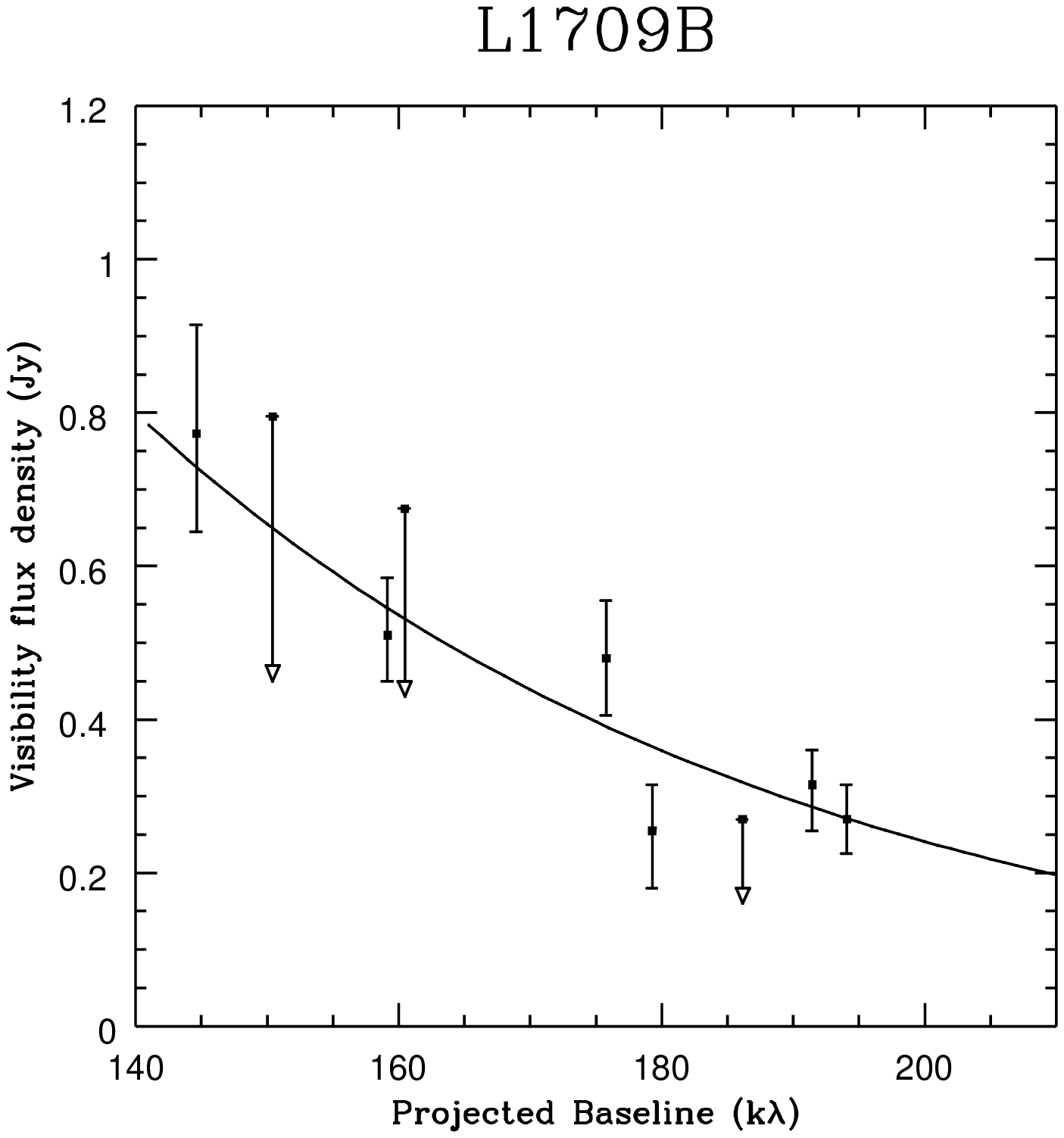}
\end{picture}
\caption{Plots of flux density versus projected baseline. (a) IRAS03282:
Note that there is no real trend in the data, and a constant flux density
of 0.4Jy is consistent with the data. This corresponds to an unresolved
disc of 0.1M$_\odot$ of radius $<$220AU. (b) L1709B:
The solid line is a gaussian fit to the data with FWHM $\sim$60AU.}
\end{figure}

\section{IRAS 03282}

IRAS 03282 is a Class 0 YSO (Andr\'e, Ward-Thompson \& Barsony 1993; 2000),
which is in a very early proto-stellar phase.
Figure 1(a) shows a plot of flux density against projected baseline for
our IRAS~03282 data. 
We note that the data are consistent with a constant
flux density of $\sim$0.4Jy
from $\sim$100--200k$\lambda$, corresponding to angular
resolution from 1.25--2.5 arcsec. At an assumed distance of 350pc (Bachiller
\& Cernicharo 1986) this corresponds to scales of 440--880AU.
We interpret this as an unresolved component, presumably a disc, around
this source. The outer radius of this disc is thus $<$220AU.

Using typical assumptions (e.g. Andr\'e et al. 2000) of 
optically thin dust emission at T=30K and mass
opacity at this frequency of $\kappa$=0.0015m$^2$kg$^{-1}$,
a flux density of 0.4Jy
at a distance of 350pc corresponds to a mass of $\sim$0.1M$_\odot$.
The single-dish submm flux density at the same wavelength is 1.4Jy in a 13
arcsec beam (Barsony et al 1998).
This corresponds to a scale of $\sim$4500AU. This is clearly emission
from the envelope around this Class 0 protostar
(see: Barsony et al 1998).
Thus the ratio of the flux in the interferometer measurements to the
single dish flux, which approximates to the ratio between the disc
emission and the envelope emission, is $\sim$30\%.

\section{L1709B}

L1709B is a Class I protostar at a distance of 140pc,
and is among
the stronger mm sources in the $\rho$ Oph dark
cloud star-forming region (Andr\'e \& Montmerle 1994).
Figure 1(b) shows a plot of flux density against projected baseline for
our L1709B data. 
In this case we see that the data are not consistent with constant
flux density on all baselines, so we have at least partially resolved
the emission.
We make the assumption that the emission is arising from an optically
thin disc, which has a gaussian radial density profile.
We plot a gaussian fit to the data and find a FWHM flux density of 0.4Jy
on a scale of 175k$\lambda$,
corresponding to a FWHM disc size of
$\sim$0.45 arcsec, or $\sim$60AU at the distance of L1709B. 
The single dish flux density of L1709B is 1.5Jy (Andr\'e et al 1993),
and we attribute this 
to the extended envelope emission around this source.
The ratio of interferometer FWHM flux to single dish flux, under typical
assumptions, is therefore $\sim$30\%.
We have also observed three other sources but can only
place limits on their fluxes. 

{\small
\begin{table}
\begin{tabular}{cllccccc}
& Source & Cloud & Class & \multicolumn{2}{c}{SBI} & JCMT & Ratio \\ 
& Name   & Name   &   &  Flux &  Scale & Flux & \\
\hspace*{0.3cm} &   &    &   &  (Jy)    & (k$\lambda$) &  (Jy) & \\ 
& & & & & & & \\ 
& L1448mm & Perseus & 0 & 0.12$^4$ & 100--200 & 1.9$^6$ & 6\% \\
& SMM3 & Serpens & 0 & 0.2$^4$ & 100--200 & 1.8$^7$ & 11\% \\
& VLA1623 & $\rho$ Oph & 0 & 0.5$^3$ & 140--200 & 4$^{8,9}$ & 12\% \\
& NGC1333IRAS2A & Perseus & 0 & 0.4$^4$ & 100--200 & 3.1$^6$ & 13\% \\
& FIRS1 & Serpens & 0 & 1.3$^4$ & 100--200 & 7$^7$ & 19\% \\
& IRAS03282 & Perseus & 0  &  0.4$^1$ & 100--200  & 1.4$^5$ & 29\% \\
& LFAM1 & $\rho$ Oph  & 0 & $\leq$1$^1$ & 140--200  & 1.8$^1$ & $\leq$55\% \\
& IRS53 & Serpens & I & 0.15$^4$ & 100--200 & 1.7$^7$ & 9\% \\
& SVS13A & Perseus & I & 0.7$^4$ & 100--200 & 2.7$^6$ & 26\% \\
& SVS20 & Perseus & I & 0.3$^4$ & 100--200 & 1.1$^6$ & 27\% \\
& L1709B & $\rho$ Oph & I & 0.4$^1$ & 175 & 1.5$^1$ & 27\% \\
& L1551-IRS5 & Taurus & I & 2.2$^2$ & 50 & 7.8$^6$ & 28\% \\
& El24 & $\rho$ Oph & II & $\leq$0.5$^1$ & 140--200 & 0.9$^1$ & $\leq$60\% \\  
& DoAr25 & $\rho$ Oph & II & $\leq$1$^1$ & 140--200 & 0.7$^1$ & $\leq$100\% \\
& HLTau  & Taurus & II & 2.6$^2$ & 50 & 2.6$^6$ & 100\% \\
\end{tabular}
\caption{{\small Comparison with previous SBI data.
Notes: $^1$This work; $^2$Lay et al 1994;
$^3$Pudritz et al 1996; $^4$Brown et al 2000; $^5$Barsony et al 1998;
$^6$Chandler \& Richer 2000; $^7$Casali et al 1993; $^8$Andr\'e et al 1993;
$^9$Ward-Thompson 1993.}}
\end{table}
}

\section{Comparison with previous SBI data}

We can compare our results with those of other workers using the same
interferometer on similar sources.
These are listed in Table 1, along with
our new detections and upper limits.
The SBI flux corresponds to the disc, and the single-dish (14'') JCMT flux is
emission from the disc-plus-envelope. 
Thus we can see a possible trend emerging from this table. 
All except one of the Class 0 sources with detections have SBI to single 
dish flux ratios of $<$20\%.
All except one of the Class I sources with detections have SBI to single 
dish flux ratios of 20--30\%.
The one detected Class II source has an SBI to single 
dish flux ratio of 100\%.
We interpret this as a
possible evolutionary trend from Class 0$\rightarrow$I$\rightarrow$II, 
caused by the ratio of the disc to envelope mass increasing
with time as a result of the envelope matter accreting onto the disc.

\section*{References}

{\small
Andr\'e \& Montmerle, 1994, ApJ, 420, 837 \\
Andr\'e, Ward-Thompson \& Barsony, 1993, ApJ, 406, 122 \\
Andr\'e, Ward-Thompson \& Barsony, 2000, `Protostars~\&~Planets~IV',~59 \\
Bachiller \& Cernicharo, 1986, A\&A, 168, 262 \\
Barsony et al 1998, ApJ, 509, 733 \\
Brown et al 2000, MNRAS, in press \\
Casali et al 1993, A\&A, 275, 195 \\
Chandler \& Richer, 2000, ApJ, 530, 851 \\
Lay et al 1994, ApJ, 434, L75 \\
Pudritz et al 1996, ApJ, 470, L123 \\
Ward-Thompson 1993, MNRAS, 265, 493
}
\end{document}